Applications of Artificial Intelligence (AI) to Network Security

Alberto Perez Veiga
University of Maryland University College

ITEC 625 - Information Systems Infrastructure

March 2018

**Abstract**

Attacks to networks are becoming more complex and sophisticated every day. Beyond the so-called script-kiddies and hacking newbies, there is a myriad of professional attackers seeking to make serious profits infiltrating in corporate networks. Either hostile governments, big corporations or mafias are constantly increasing their resources and skills in cybercrime in order to spy, steal or cause damage more effectively. With the ability and resources of hackers growing, the traditional approaches to Network Security seem to start hitting their limits and it's being recognized the need for a smarter approach to threat detections.

This paper provides an introduction on the need for evolution of Cyber Security techniques and how Artificial Intelligence (AI) could be of application to help solving some of the problems. It provides also, a high-level overview of some state of the art AI Network Security techniques, to finish analysing what is the foreseeable future of the application of AI to Network Security.

## Introduction

To add up to the fact that cyber-attacks are alarmingly growing in amount and complexity, one of the scariest facts about Cybersecurity for companies and organizations is the lack of readiness, mostly from a business perspective. The problem is wider than the technical gap. Management layer lack awareness and understanding of the real needs, therefore not providing the required support. This lack of support causes for many organizations the subsequent lack of traction, attention and willingness to commit funding and resources to Cybersecurity. It's especially important to mention the lack of properly trained personnel to fill all the short-term future needs for Cybersecurity positions. If this tendency continues, in 2021 approximately there 3.5 million positions needed related with Cybersecurity will be unfilled and cybercrime may cost a total of $3 trillion around the world (Morgan, 2017).

Given the current status, it's easy to realize why Cybersecurity experts have been seriously looking into Artificial Intelligence (AI) and how it could help alleviate some of these issues. As an example, Machine Learning (ML), used by many of the most recent AI algorithms can greatly help with the detection of malware, increasingly difficult to identify and isolate. With malicious software becoming more capable of adapting to linear traditional security solutions, ML provides the ability to learn not only how malware looks like and acts, but also how it may evolve. AI systems could help, additionally, not only providing detection, but also take proactive actions that could take steps to remediate certain situations, and also sort and categorize events and threats, eventually freeing up technicians from repetitive, activities (Kh, 2017).

Some studies estimate that the need for reliable data will boost the investment in big data analytics and intelligence up to $96 million by 2021. Sectors in need of the best security solutions such as defence or banking are pushing these technologies, adopting them and

investing. Established antimalware companies such as Symantec are pushing the boundaries of their technologies and started embedding AI algorithms in their solutions (Abi Research, 2017). Other companies, however, have jumped directly into AI to develop security products.

## Machine Learning

The first that has to be mentioned is that, beyond all the buzz of many companies claiming to be using AI in their Cybersecurity appliances, AI is not fully applied, for now, in the field of Cybersecurity. To be fair, the current applications of AI are mostly restricted to Machine Learning (ML). While we could consider AI to have the ultimate objective to make machines function with some sort of intelligence or, in other words, being what we consider "smart", ML is a subfield of AI which studies the way in which computers can learn the better way to perform their intended function without the need of being explicitly programmed to perform such functions. ML may encompass techniques such as statistics, mathematical optimizations, or data mining (Crosby, 2017). ML algorithms try to make decisions about their behaviour and find ways to solve problems by inferring them from models based on sample inputs that represent real-life scenarios.

One of the main challenges for Network Security consists in the fact that attacks change often their appearance and vectors. When a new kind of attack or malware strikes, is not possible for traditional system to detect and identify these behaviours, as there are no stablished rules or previous patterns against which it can be matched. A typical example of this situation are *zero-day attacks*. When an attacker uses a zero-day, exploits certain vulnerability that has been discovered by someone, but not yet known or patched by the vendor. These undiscovered vulnerabilities are so effective that they have a huge demand in criminal forums, being sold in black markets in the dark web, some of them for large amounts of money.

The approach from traditional systems is to stop malware before it executes, trying to match patterns of code with known signatures. However, when they fail to do it, there is usually little to no remediation left. The malware executes and it's very difficult to stop. ML algorithms try to identify in real-time when the malware has just stroke and, together with AI-assisted decisions, machine and network isolation techniques, isolate the infected computer or whole network segments in milliseconds, providing a way to stop the spread of the malicious code.

In any case, when applying AI or ML to network security, one of the most relevant problems still seems to be how to identify useful patterns and how to understand when a deviation from them constitutes a security event, classify it and act upon it accordingly. Statistic deviations from normal network traffic may result interesting from a "normality" perspective, however, not all of them constitute necessarily security incidents. A negative detection can be as problematic as a huge number of false positives.

**Supervised ML**

Supervised ML tries to reach the state where the computer learns a function to map some input to an output based on samples data which provides such mapping examples from real-life scenarios. In general, these algorithms try to infer this function from the sampled data, creating a function that will map inputs (x) to outputs (Y) in the form Y=f(x). The function is supposed to become precise enough to calculate new outputs accurately from new inputs. Supervised problems can be grouped into two main groups: Classification and Regression problems. The first one, classification is where the output of the function is a classification, e.g. a colour: Red, blue, etc. In the case of Regression problems, the output is a value such as a measure, weight, cost, etc.

Unfortunately, there are some areas, such as the detection of attacks based on network traffic, of direct application to Intrusion Detection Systems (IDS) and Intrusion Prevention Systems (IPS), where the lack of sample data is a problem. During the last two decades, it's been proven very difficult to find comprehensive sets of data samples. Although some good efforts have been made in order to sample network traffic and provide analysis, such as MIT datasets provided for years 1998, 1999 and 2000 (MIT, n.d.), these samples have been found to be irrelevant or biased (McHugh, 2000).

Despite the caveats, supervised ML is probably the field of AI which has delivered, until now, the best results for Cybersecurity. The immense databases containing malware files, allow to train these algorithms deeply, improving accurate detection and reducing the number of false positives and negatives. The same applies to spam detection, which benefits from huge samples databases, constantly expanding with the help of users and Internet Services Providers reporting spam messages. ML provides a smarter, more effective approach, rather than the traditional text-based string-matching or address blacklisting. Classification algorithms can separate between "spam" and "legitimate" emails by inferring the classification from the presence of intentionally misspelled words (V!4GRA, S11mming P1lls, etc…), or the presence in the email body of URL links to sites delivering malware or of doubtful reputation (Kanal, 2017).

Another example of the application of Supervised ML to Network Security would be the use of Recurrent Neural Networks (RRN) to separate human-generated DNS records from those automatically generated by ransomware. Blacklisting techniques based on signatures are completely useless for this case, as malicious names keep constantly growing in number, changing and evolving. However, ML with RRN can be used to achieve more accurate results by using linguistic analysis, thanks to the fact that malware-generated domains tend to have strange vowel/consonants ratios that can be identified and isolated using RRNs (Machine Learning…, 2017)

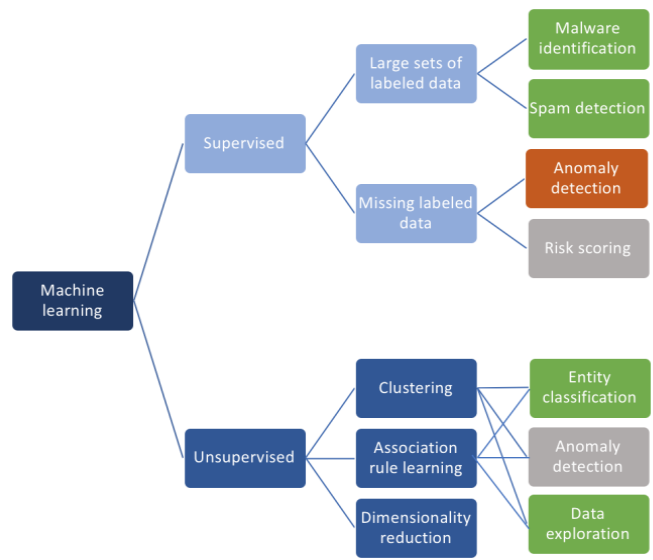

Figure 1, graph depicting a partial view of Machine Learning algorithms and their applications to Cybersecurity (Marty, 2018)

**Unsupervised ML**

Despite the fact that Supervised ML has probably delivered the best results until now, it's important to mention that due to its inherent limitations, the focus in research when it comes to Network Security seems to be shifting towards Unsupervised ML. For Unsupervised ML, no sample labelled data is provided to the ML algorithm. The computer, in this case, is expected to learn the underlying structure from the data and ultimately infer proper outputs. Unsupervised ML is much more subjective and patterns obtained don't often make sense without the human expertise to make sense of them and understand which of those patterns are really useful. Some interesting techniques with application to Cybersecurity are: Clustering, Association Rule Learning and Dimensionality Reduction.

Clustering techniques try to group objects that are similar in clusters, which can be used to find out i.e. if there is an abnormal amount of network traffic in a certain host, a significant number or wrong login events, a user accessing data which is normally not accessed by him, users working out of their usual working hours, connections from unusual locations, etc. Association Rule Learning techniques try to discover the relationships and their rules between elements in large databases.

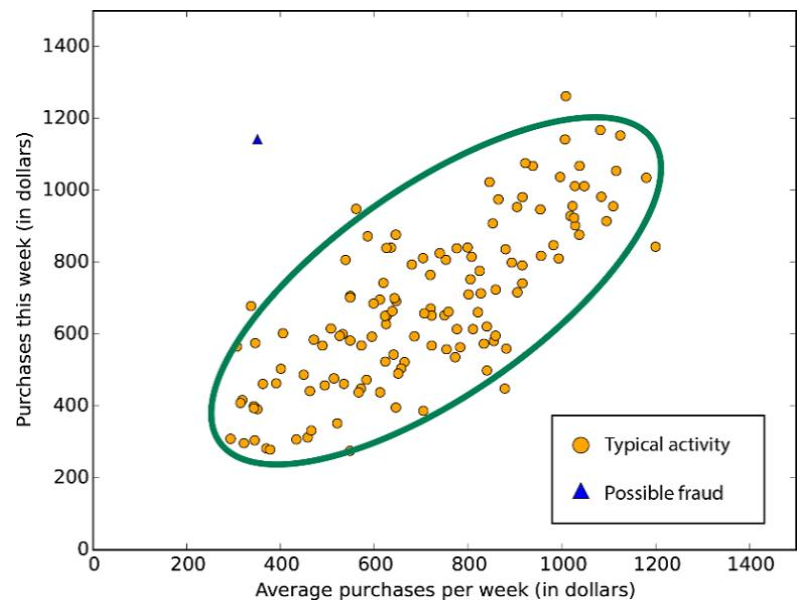

Fig. 2 Data Clustering applied to the detection of online fraud (Microsoft, 2017)

Dimensionality Reduction techniques may be particularly helpful for Network Security. The whole idea behind them is to find the way to reduce the number of *features* in a set of data that may be of application to solve a certain problem. A *feature* is, in general, a particular attribute of the different elements in a set of data. Let's take as an example a capture of network traffic. In this particular case, there may be multiple *features*: origin IP, destination IP, protocol, port, payload, MAC addresses, TTL, routing, etc. Analysing all the features of the captured data might be computationally ineffective to solve a particular security problem. The problem could become completely unsolvable if we are dealing with real-time analysis. If the ML algorithm can, however, learn how to reduce the number of relevant features to a few (Feature selection techniques), or group them into more manageable sets (Feature Extraction) to solve the problem, that would make it easier to be solved.

Self-Organising Maps (SOMs) are a type of Neural Network in which each neuron is connected to its neighbours and represents a point in a multidimensional space, although they are represented in a bi-dimensional space. SOMs, together with clustering techniques can help identifying malicious IP traffic. When feeding the system with sample data, the closest neuron to normal traffic moves towards it, "pulling" the neighbour neurons with it.  When feeding real data, the neurons will cluster around normal data, while anomalous data will do the opposite and be isolated or separated, helping identifying patterns of anomalies, which probably constitute malicious traffic (Machine learning…, 2017). Figure 3 depicts an example of a SOM constituted by the automated analysis of more than 3.000 applications to the University of Finland. No parsers, classifications or taxonomies were used to achieve it, but only the analysis of the documents.

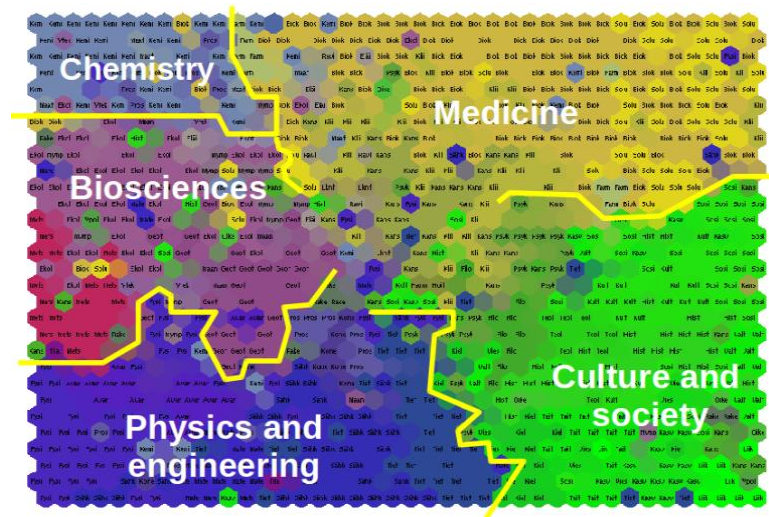

Fig. 3. Example of Self Organizing Map (Honkela, 2011)

It's interesting to mention how Data Science can be combined with unsupervised ML in order to achieve more accurate results. An example of this is the application of Hidden Markov Models (HMMs). These models have been traditionally applied to fields such as speech recognition or biological analysis. A Markov Model is essentially a mathematical a model that produces future data based on the values of present data, rather than past states. This model is of application e.g. to weather or economic predictions. A Hidden Markov Model is nothing else than a model which generates an educated guess. Based on the knowledge about Markov Models, which we can use it to deduct the underlying Markov Model (Alghamdi, 2016). HMMs are, therefore, of application when analysing data in an unsupervised ML environment, where there is no network traffic sample data with clear states and outputs from which infer a Markov Model.

## Practical case - Darktrace

Different companies are applying these techniques in order to increase the effectivity of their solutions. There are other companies such as Darktrace, which are heavily investing in AI and ML in order to achieve spectacular results fighting against emerging threats.

Darktrace is a technology company founded in the UK in 2013 by experts in AI from the University of Cambridge, together with experts from different government intelligence Agencies. They base their whole business model in security solutions fully developed on the top of AI and Machine Learning techniques. Darktrace's main marketing point consists in the way they use AI within their products to detect and defend against malicious activity: they claim to have patented an algorithm known as "Enterprise Immune System", capable of defending an enterprise network emulating the way the human body defends against infections: being able of differentiate "good" cells from "bad" cells, attacking and isolating the bad cells in the same manner as human antibodies work. Human body is capable of understand, in most of the cases, when a virus or a bacteria is causing any harm, as soon as it starts causing it.

Darktrace claims to be able to catch network attacks without the use of rules, using Machine Learning techniques that grant real-time detection. They benefit particularly from Unsupervised ML techniques, which grant the ability to catch unknown threats due to the fact that the learning is not based on known datasets, rules, or models, allowing the computer to self-learn patterns of normality and abnormal behavior.

Some of the successful use cases claimed by Darktrace include i.e. the detection of a new strain of complex Ransomware attacks. In this particular case, an employee accessed its personal email from the corporate network, allowing the malicious software to enter the network and start accessing SMB shares and encrypting them. This behavior was understood as malicious by the engine, which flagged it as a threat nine seconds later. As this happened during the weekend, the security team was gone. 24 seconds later the engine took autonomously the decision to stop the anomalous writing activities to the SMB drives, stopping the attack before it was able to spread further, neutralizing the attack and limiting the damage to a small amount of data.

Another interesting example could be the detection of the exfiltration of information carried on by a disgruntled employee. The engine was able to detect that a large amount of data was being transferred to an external cloud provider. Given the fact that the server from which it was being transferred was not usually carrying on those types of transactions, this was considered as an unusual behavior and flagged it as a possible threat and reported. The information was found to be thousands of records containing personal data from clients of the company being exfiltrated by a disgruntled employee.

The security of the Internet of Things (IoT) has been proven to be a constant nightmare, due to the fact that most of these devices embed little to no security mechanisms at all. A good example of how these engines can help with this problem is how a biometric scanner was compromised and an attacker started changing data exploiting telnet connections from an external computer and then pivoting to other servers in the network to eventually gain more privileges. Given the fact that the engine had learned about the "normal profile" of the device, understood that this kind of activity was deviating from it and flagged it as an alert. Investigation revealed that the vulnerability and data of these Biometric Scanners were exposed in the search engine *Shodan* and actively exploited for malicious purposes.

All these examples highlight cases for which traditional approaches will be most likely completely ineffective. The detection of Ransomware has been proven to be a nightmare for most traditional Anti-Virus (AV) solutions, due to its constant shifting in the way they operate, the vectors used and even encryption techniques. Activities such as data thief carried on by internal employees are also very difficult to detect, as the employee is not usually carrying an activity that can be easily flagged as malicious: the employee usually has the access rights to the data. Besides these application cases, there are many other in which unsupervised ML could definitely help.

**The foreseeable future of AI in Cyber Security**

Despite all the significant advances that the world of Network Security has experienced during the last years, especially in terms of the adoption of AI, it seems wise to be cautious about the range of applications of it. It's a temptation to believe that AI is the silver bullet that can solve all the Cyber Security problems, or blindly believing that it will success in every pitfall where traditional techniques have failed.

Behind all the buzz behind the words "Artificial Intelligence", we should be clearly aware that we are not there yet. There are, for now, only certain techniques that are actually being used with good results in security applications, and these are constrained to Machine Learning. Although systems are far from being "Intelligent" and aware of their own level of knowledge, as it's expected from AI, Supervised ML has delivered some interesting results until now and Unsupervised ML seems to be the focusing most of the research in this discipline, as well as delivering some spectacular results and achieve levels of accuracy not possible only a few years ago. However, Unsupervised ML is still highly dependent on human expertise to have context and knowledge to make sense of data.

It seems, therefore, that the future steps in AI will be to keep exploring the path of Unsupervised ML and keep working to try and eliminate as much as possible the need for human interaction. It's important to keep working towards finding algorithms capable to understand the context where they are operating. I.e. understanding that the traffic generated by a DNS server is the way it is *because* it's a DNS server, rather than having to blindly build a "normality profile" that would flag as malicious traffic something as usual as the federation with another DNS server to delegate zones. In other words, algorithms need to evolve to understand why there is a particular pattern behind a particular behaviour, rather than learning and assuming it blindly.

Additionally, the next step would be to build algorithms capable of providing expert knowledge, rather than needing humans in order to identify i.e. the patterns discovered by SOMs or clustering techniques. One of the concepts that seems to be gaining traction in this direction is the utilization of Bayesian Belief Networks (BNs) in order to create expert systems. BNs, also known as Causal Probabilistic networks are a way to represent relationships among different events in terms of probability.

Finally, it's important to keep working in the visualization of data to improve the ability of security analysts to understand a broader range of threats, in less time with less effort and resources committed to it. With the booming of Big Data, the ability to store and analyse immense amounts of data is becoming a pressing need where ML is playing an important role. However, it's not only important to be able to analyse it, but also to present it in a format that can be easily understood at multiple levels in an organization. Data visualization is, therefore, one of the aspects where ML will most likely play a prominent role in the future.

## Conclusion

The IT market has, a tendency to quickly assimilate buzz words pushed by marketing departments. During the last years, technologies such as Big Data, Cloud Computing, Artificial Intelligence, etc., have been repeated again and again in multiple forums, in many cases without a clear understanding of their significance or their application to solving real problems effectively. It's a known fact that when humans don't completely understand a technology, two kinds of effects usually occur: either the technology is irrationally rejected (e.g. new operating systems) or, if it's properly marketed, it's assumed to be the silver bullet capable of solving every problem (Artificial Intelligence). It usually takes some time, even years, for the dust to settle down and for the market to realize the true potential of it.

The irruption of ML in in Cyber Security is forcing a paradigm shift from proactive rule-based prevention, to reactive real-time detection. Security threats have become so varied, different and smart, that traditional techniques, based on rules inferred from known attacks, that stop the attack before it happens, don't seem to be a viable approach anymore. Many attacks escape these mechanisms and cause tremendous damage that can't be stopped once it has started. ML aims at identifying attacks in real-time, with little to no human interaction and stopping them before they provoke serious harm.

We can conclude that Artificial Intelligence is not being used nowadays, as it's expected to be, to solve Network or Cyber Security problems in general. For now, only Machine Learning, a branch of AI, is being successfully applied to solve a small part of the problems. Supervised ML has delivered a number of interesting practical solutions, however, there is ongoing research, particularly towards the utilization of Unsupervised ML, as the ultimate goal is to reduce human interaction as much as possible when detecting threats. The only viable way to evolve Security and get AI closer to what it's expected to deliver is to keep investigating and find new techniques capable of providing context, expertise and enhanced data visualization, as well as achieving a tighter integration with Data Science techniques and ML-enhanced data analytics.

Applications of Artificial Intelligence (AI) to Network Security                17Marty, R. (2018, January 11). AI in Cybersecurity: Where We Stand & Where We Need to Go. Retrieved March 12, 2018, from https://www.darkreading.com/threat-intelligence/ai-in-cybersecurity-where-we-stand-and-where-we-need-to-go/a/d-id/1330787

McHugh, J. (2000). Testing Intrusion detection systems: A critique of the 1998 and 1999 DARPA intrusion detection system evaluations as performed by Lincoln Laboratory. ACM Transactions on Information and System Security (TISSEC), 3(4), 262-294. Retrieved March 12, 2018.

Microsoft. (2017, December 18). How to choose machine learning algorithms. Retrieved March 24, 2018, from https://docs.microsoft.com/en-us/azure/machine-learning/studio/algorithm-choice

MIT. (n.d.). DARPA Intrusion Detection Data Sets. Retrieved March 17, 2018, from https://www.ll.mit.edu/ideval/data/

Morgan, S. (2017, June 06). Cybersecurity labor crunch to hit 3.5 million unfilled jobs by 2021. Retrieved March 10, 2018, from https://www.csoonline.com/article/3200024/security/cybersecurity-labor-crunch-to-hit-35-million-unfilled-jobs-by-2021.html

Kh, R. (2017, December 01). How AI is the Future of Cybersecurity. Retrieved March 10, 2018, from https://www.infosecurity-magazine.com/next-gen-infosec/ai-future-cybersecurity/